\documentclass[prl,twocolumn,showpacs,superscriptaddress]{revtex4-1}
\usepackage{pdfpages}
\usepackage{amsmath,amssymb,braket,cases}
\usepackage{graphicx}
\usepackage{booktabs,multirow}
\usepackage{hyperref}
\usepackage{float}
\usepackage{color}
\usepackage{dcolumn}
\usepackage{bm}
\usepackage[flushleft]{threeparttable}
\usepackage{multirow}
\usepackage{array}
\usepackage{url}
\usepackage{booktabs}
\usepackage{mathtools}

\usepackage{color}
\hypersetup{colorlinks=true, urlcolor=blue, citecolor=cyan, pdfborder={0 0 0},}

\newcommand{\be}{\begin{equation}}
\newcommand{\ee}{\end{equation}}
\newcommand{\beq}{\begin{eqnarray}}
\newcommand{\eeq}{\end{eqnarray}}
\newcommand{\bk}{{\bf k}}

\newcolumntype{P}[1]{>{\centering\arraybackslash}p{#1}}
\usepackage[all]{hypcap}
\makeatletter
\AtBeginDocument{\let\LS@rot\@undefined}
\makeatother

\begin{document}

\title{The Dirac-Weyl semimetal: Coexistence of Dirac and Weyl fermions in polar hexagonal $ABC$ crystals}

\author{Heng Gao}
\affiliation{International Centre for Quantum and Molecular Structures, Department of Physics, Shanghai University, 99 Shangda Road, Shanghai 200444, China}
\affiliation{Department of Chemistry, University of Pennsylvania, Philadelphia, Pennsylvania 19104--6323, USA}
\author{Youngkuk Kim}
\affiliation{Department of Chemistry, University of Pennsylvania, Philadelphia, Pennsylvania 19104--6323, USA}
\affiliation{Department of Physics, Sungkyunkwan University, Suwon 440-746, Korea}
\author{J\"orn W. F. Venderbos}
\affiliation{Department of Chemistry, University of Pennsylvania, Philadelphia, Pennsylvania 19104--6323, USA}
\affiliation{Department of Physics and Astronomy, University of Pennsylvania, Philadelphia, Pennsylvania 19104--6396, USA}
\author{\begin{tabular}{cc} C. L. Kane\end{tabular}}
\affiliation{Department of Physics and Astronomy, University of Pennsylvania, Philadelphia, Pennsylvania 19104--6396, USA}
\author{E. J. Mele}
\affiliation{Department of Physics and Astronomy, University of Pennsylvania, Philadelphia, Pennsylvania 19104--6396, USA}
\author{Andrew M. Rappe}
\email{rappe@sas.upenn.edu}
\affiliation{Department of Chemistry, University of
Pennsylvania, Philadelphia, Pennsylvania 19104--6323, USA}
\author{Wei Ren}
\email{renwei@shu.edu.cn}
\affiliation{International Centre for Quantum and Molecular Structures, Department of Physics, Shanghai University, 99 Shangda Road, Shanghai 200444, China}

\date{\today}

\begin{abstract}
We propose that the noncentrosymmetric LiGaGe-type hexagonal $ABC$ crystal SrHgPb realizes a new type of topological semimetal that hosts both Dirac and Weyl points in momentum space. The symmetry-protected Dirac points arise due to a band inversion and are located on the sixfold rotation $z$-axis, whereas the six pairs of Weyl points related by sixfold symmetry are located on the perpendicular $k_z=0$ plane. By studying the electronic structure as a function of the buckling of the HgPb layer, which is the origin of inversion symmetry breaking, we establish that the coexistence of Dirac and Weyl fermions defines a phase separating two topologically distinct Dirac semimetals. These two Dirac semimetals are distinguished by the $\mathbb{Z}_2$ index of the $k_z=0$ plane and the corresponding presence or absence of 2D Dirac fermions on side surfaces. We formalize our first-principles calculations by deriving and studying a low-energy model Hamiltonian describing the Dirac-Weyl semimetal phase. We conclude by proposing several other materials in the non-centrosymmetric $ABC $ material class, in particular SrHgSn and CaHgSn, as candidates for realizing the Dirac-Weyl semimetal.
\end{abstract}

\maketitle
\setlength{\parindent}{4ex}
\setlength{\parskip}{0em}

{\it Introduction.}---Since the discovery of topological insulators \cite{Hasan10p3045, Qi11p1057}, the classification of materials based on symmetry and topology has been greatly extended, encompassesing both insulating states and semimetals. For instance, a rich variety of topological phases distinct from time-reversal invariant topological insulator have been predicted and discovered, such as topological crystalline insulators \cite{Fu11p106802, Alexandradinata14p116403,Tanaka12p800} and topological superconductors \cite{Qi09p187001,Sato17p076501}. The classification of electronic band structures based on symmetry and topology can be extended to gapless systems, giving rise to distinct types of (semi)metallic phases with protected nodal degeneracies \cite{Armitage18p015001}. Examples include Weyl semimetals \cite{Wan11p205101}, Dirac semimetals \cite{Young12p140405, Liu14p864, Liu14p677,Steinberg14p036403}, line-node semimetals \cite{Kim15p036806, Yu15p036807}, and double Dirac semimetals \cite{Wieder16p186402,Bradlyn16p5037}.\par

Weyl semimetals are characterized by isolated point touchings of two non-degenerate bands in momentum space. The {\em twofold} degenerate nodal points, called Weyl points/nodes, correspond to quantized monopoles of Berry curvature and cannot be removed unless two nodes of opposite monopole charge come together and annihilate. The monopole charge, referred to as the chirality of the Weyl node, is given by a topological invariant in momentum space (i.e., the Chern number) and determines the dispersion away from nodal point. Weyl semimetals can be considered symmetry-prevented topological phases; since the presence of both time-reversal ($T$) and inversion ($P$) symmetry forces all bands to be twofold degenerate, Weyl points can only occur when at least one of these symmetries is broken \cite{Vafek14p83}.\par

In contrast, bulk Dirac points are {\em fourfold} degenerate nodal band touchings with linear dispersion and vanishing net chirality, consistent with $T$ and $P$ symmetry. Their stability, however, requires additional crystalline symmetries, since without additional symmetries the two degenerate nodes of opposite chirality forming the Dirac point can annihilate or separate \cite{Young12p140405,Yang14p4898,Gao16p205109}. Therefore, Dirac semimetals are examples of symmetry-protected topological (gapless) phases. Dirac semimetals protected by rotation symmetry have been predicted and experimentally observed in Na$_{3}$Bi and Cd$_{3}$As$_{2}$  \cite{Liu14p864, Liu14p677}. These semimetallic phases can exhibit a rich set of topological phenomena, such as tunable transitions to distinct gapped phases with different topology \cite{Young12p140405}, and the chiral anomaly induced by an external magnetic field \cite{Gorbar14p085126}. Due to the different---and seemingly contradictory---symmetry requirements of Weyl and Dirac semimetals, one may wonder whether these phases can coexist in real materials. \par

Motivated by this question, in this Letter we predict that noncentrosymmetric SrHgPb with hexagonal $ABC$-type crystal structure realizes a new topological semimetal phase with coexisting Dirac and Weyl points: the Dirac-Weyl semimetal. Based on first-principles calculations, we show that SrHgPb with polar space group $P6_{3}mc$ (\# 186) hosts a pair of Dirac fermions along the polar rotation axis and six pairs of Weyl fermions in the plane perpendicular to the polar axis. The Dirac points originate from a band inversion and are protected by crystal point group symmetry, in particular the sixfold rotation symmetry. The local stability of the Weyl points in the $k_z=0$ plane is  guaranteed by a combined twofold rotation and time-reversal symmetry; the existence and location of the Weyl points can be controlled by the buckling of the HgPb layers. In particular, we demonstrate that, as a function of the buckling, Weyl points of opposite chirality can be paired and annihilated, marking the transition between phases for which the $k_z=0$ plane has nontrivial/trivial $\mathbb{Z}_2$ index. For a buckling which minimizes the total energy, each pair of Weyl points is well separated by a distance 0.07 $\text{\AA}^{-1}$, suggesting that SrHgPb should be a promising material candidate for studying the Fermi arc surface states experimentally. To describe the Weyl-Dirac semimetal phase, we develop a simple low-energy model for a band inversion of spin-$\frac32$ and spin-$\frac12$ doublets in polar materials. We conclude by examining other members of the hexagonal $ABC$-type material class, which leads us to a prediction of further candidate materials realizing coexisting Dirac and Weyl nodes.\par

\begin{figure}[pt]
    \centering
    \includegraphics[width=8.6cm]{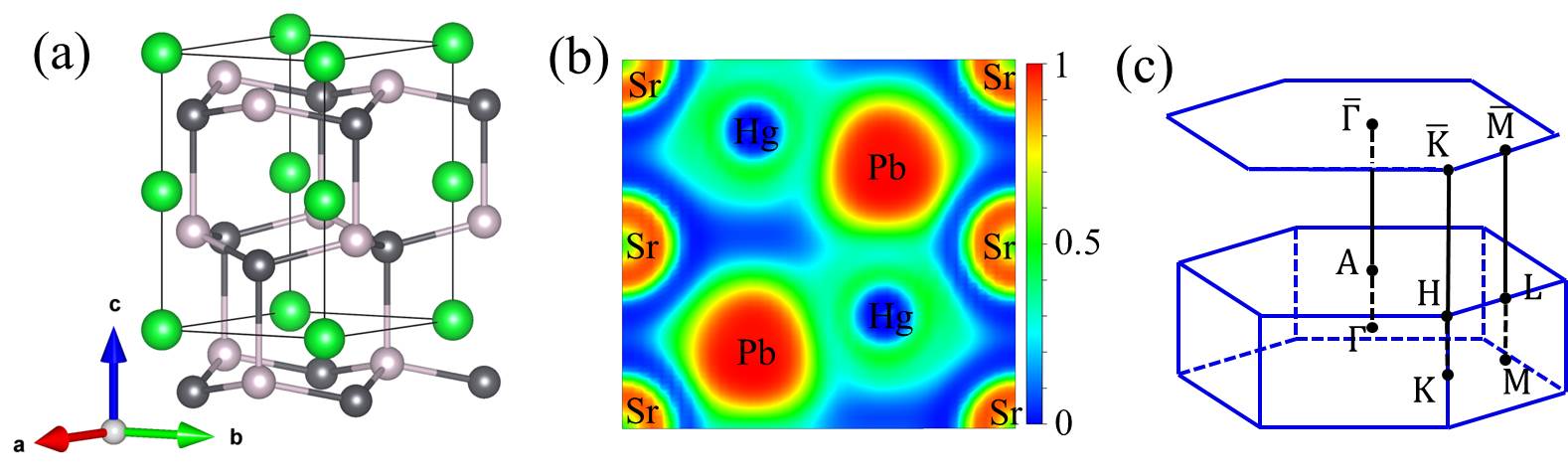}
    \caption{(a) The crystal structure of SrHgPb in the $P6_{3}mc$ space group. The green, black and gray
    spheres indicate Sr, Pb, and Hg atoms, respectively. (b) The electron localization function
    of SrHgPb on the (110) plane. (c) The first Brillouin zone and its projection onto the (001) surface.}
    \label{fig:fig_1}
\end{figure}

{\it Crystal structure of SrHgPb.}---The ternary compound SrHgPb has been synthesized \cite{Merlo93p145} in a hexagonal structure of LiGaGe-type, also known as a stuffed wurtzite lattice [shown in Fig.\,\hyperref[{fig:fig_1}]{1(a)}]. The generating point group of $P6_{3}mc$ is isomorphic to $C_{6v} $ and contains elements which require a half translation along the $z$-direction, such as the sixfold screw rotation $S_{6z}$. The proper $C_{3v}$ subgroup which does not require fractional translations is generated by the threefold rotation $C_{3z}$ and mirror reflection $M_{yz}$. The unit cell of SrHgPb consists of two buckled HgPb layers and two Sr atoms occupying the interstitial sites of the HgPb wurtzite lattice. The buckling of the HgPb layer is calculated to be 0.78 $\text{\AA}$ here, in line with the experimental result \cite{Merlo93p145}.  To elucidate the bonding character of SrHgPb, we calculate the electron localization function (ELF) \cite{Silvi94p683} on the (110) plane that contains the Sr nuclei, as shown in Fig.\,\hyperref[{fig:fig_1}]{1(b)}. The calculated ELF shows that the Pb atoms form covalent bonds with four nearest-neighbor Hg atoms, while the Sr atoms form ionic bonds with the HgPb wurtzite lattice. We also calculate the Bader charges \cite{Graeme06p354} of SrHgPb to obtain the charge transfer between the Sr atom and the HgPb wurtzite lattice, further confirming the mixed bonding nature of SrHgPb. Sr transfers 0.5 $e$ to Hg and 0.8 $e$ to Pb, in good agreement with the results of the ELF calculations.\par

\begin{figure}[pt]
    \centering
    \includegraphics[width=8.6cm]{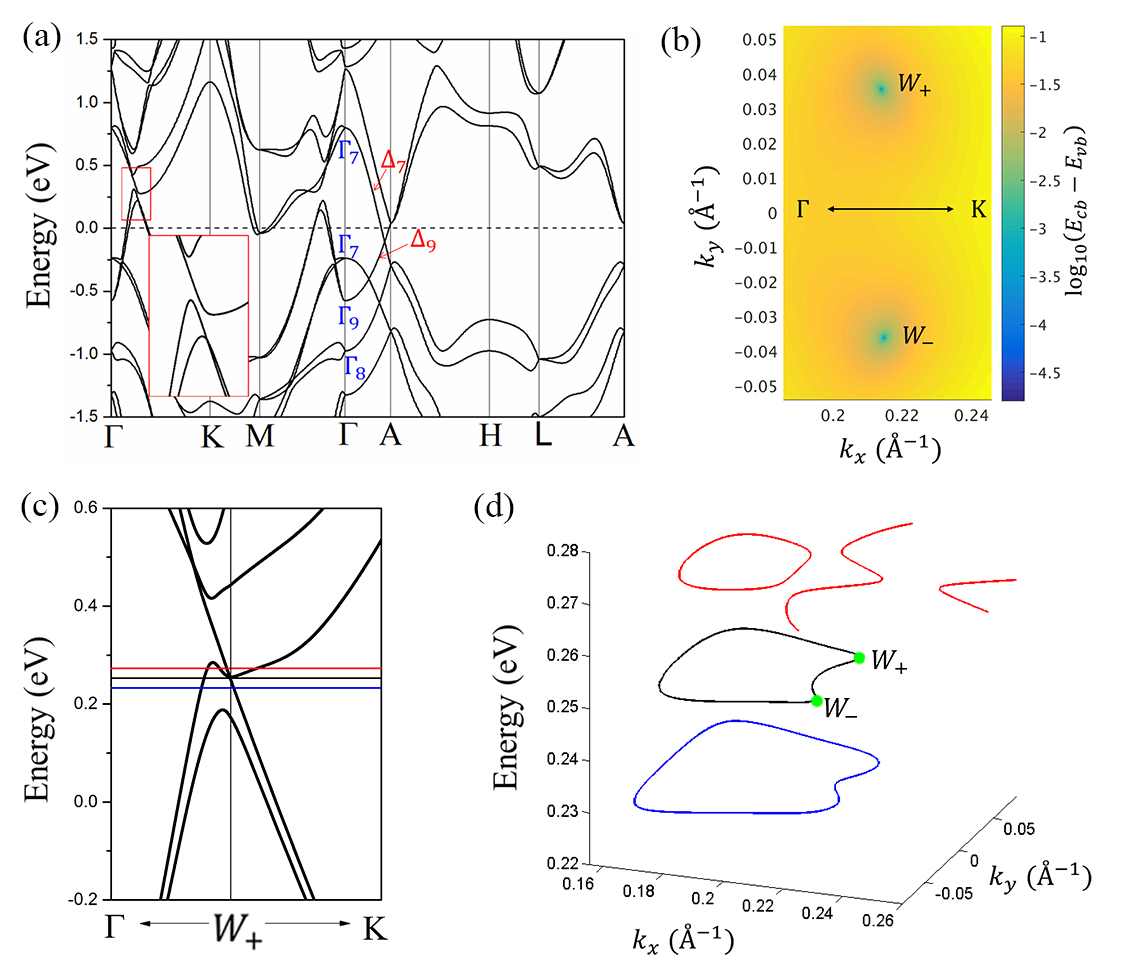}
    \caption{The electronic structure of SrHgPb. (a) Band structure along the high-symmetry lines of  momentum space. The inset provides a magnified view of the highlighted box region. (b) Color map of energy difference between the conduction and valence bands in logarithmic scale $\log_{10}(E_{cb}-E_{vb})$, where    $W_{+}$ and $W_{-}$
    denote the Weyl points. (c) The band structure along $\Gamma$-$W_{+}$-K. $W_{+}$ is an off-symmetry point at (0.217, 0.036, 0) $\text{\AA}^{-1}$, where the valence band
    and conduction band touch and form a Weyl point. (d) The Fermi surfaces in the $k_{z}=0$ plane, for $E_f$ energy levels corresponding to the red, black, and blue lines in (c). The two green dots indicate Weyl points.}
    \label{fig:fig_2}
\end{figure}

{\it Symmetry-protected Dirac points.}---The relativistic band structure of SrHgPb obtained from first-principles calculations is shown in Fig.\,\hyperref[{fig:fig_2}]{2(a)}. The band structure along high symmetry lines shows that SrHgPb is a metal, with both electron and hole pockets located in the vicinity of $\Gamma$ (hole), $M$ (electron), and $A$ (electron). Generically, energy bands are non-degenerate in non-centrosymmetric systems, except for points, lines or planes of high symmetry. In the present case, the twofold screw rotation $S_{2z}$ mandates a twofold degeneracy of energy bands in the $k_z=\pi$ plane, since $S^2_{2z}=-e^{ik_z}$ and thus $(S_{2z}T)^2=-1$. Furthermore, the anti-commutation relation $\{S_{2z},M_{yz} \} = 0$ requires twofold degeneracies along the high symmetry lines $ \Gamma\text{\textendash}A$ and $M \text{\textendash}L$. In particular, the crossing of conduction and valence bands along $ \Gamma\text{\textendash}A$, labeled by $\Delta_7$ and $\Delta_9$ in Fig.\,\hyperref[{fig:fig_2}]{2(a)}, is a crossing of twofold degenerate bands, forming a pair of four-fold degenerate Dirac points located at Brillouin zone coordinates (0, 0,$\pm$0.33) $\rm{\AA}^{-1}$ and energy $E - E_{f} = -0.1$ eV. Since the bands which cross have different symmetry, the Dirac points are symmetry protected. 

The origin of the Dirac points lies in a band inversion at $\Gamma$, resulting from an energetic splitting of bonding and anti-bonding states of the two $B$ and $C$ sites in the unit cell. In this sense, the Dirac points are similar to the known centrosymmetric Dirac materials Na$_{3}$Bi \cite{Liu14p864}, but should be distinguished from semimetals with essential fourfold degenerate Dirac points, mandated by crystal symmetry and located at high-symmetry points, such as BiO$_{2}$ \cite{Young12p140405}. Similar noncentrosymmetric Dirac semimetals, arising due to band inversion, have been proposed in YbAuSb \cite{Gibson15p205128} and LiZnBi \cite{Cao17p115203} with LiGaGe-type structure.\par

\begin{figure}[pt]
    \centering
    \includegraphics[width=8cm]{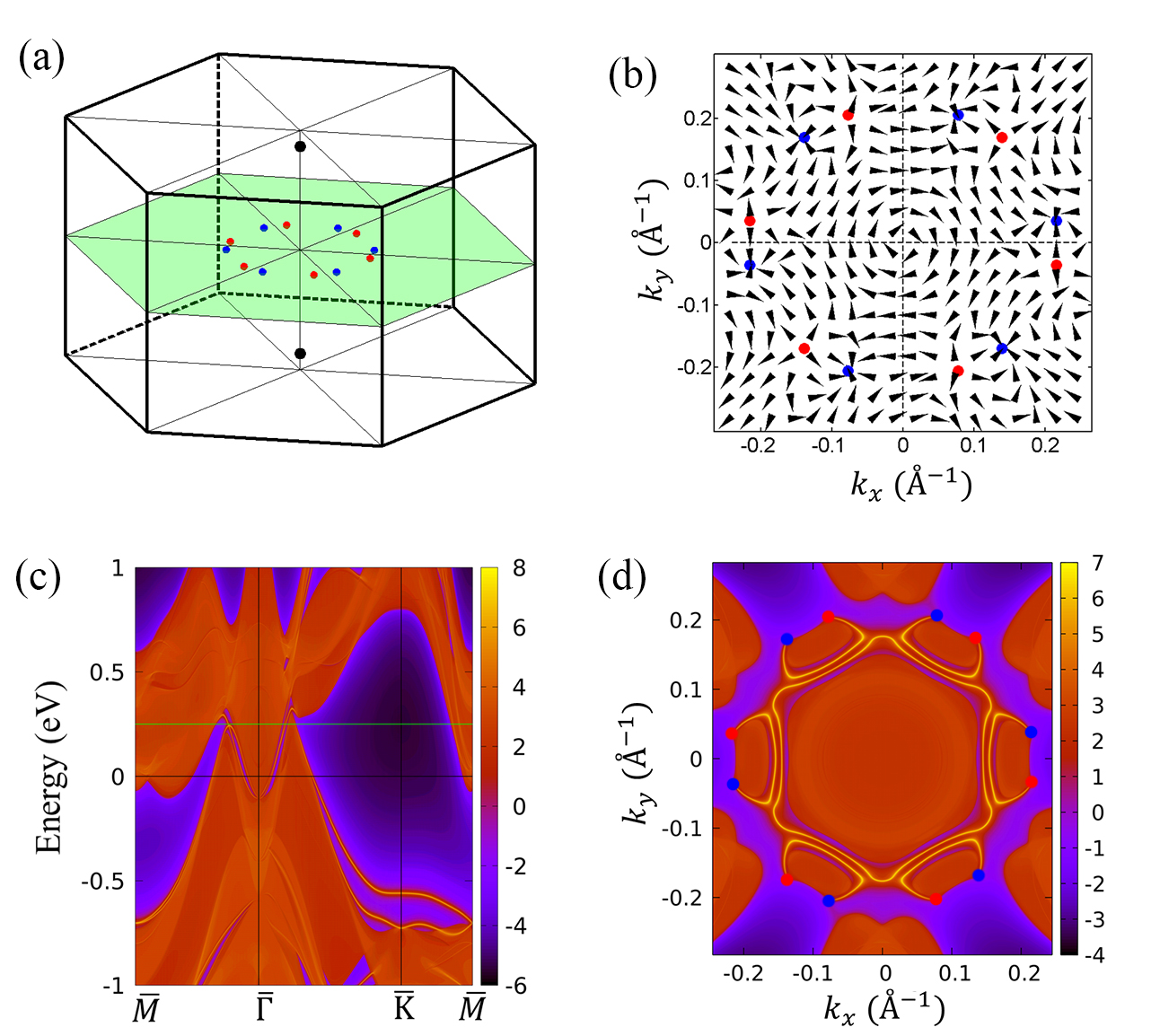}
    \caption{(a) The distribution of the Dirac points (black dots) and the Weyl points (blue and red dots) in the Brillouin zone. (b) In-plane Berry curvature on $k_z$=0
    plane, including six pairs of the Weyl points. The arrows denote the direction of the in-plane Berry curvature. (c) The surface band structure of the (001) surface. (d)
    The density of surface states at $E - E_{f} = +0.25\,\rm{eV}$ indicated by a green line in (c). The red and blue points illustrate the projective points of the bulk Weyl points
    with +1 and -1 Chern numbers.}
    \label{fig:fig_3}
\end{figure}

{\it Type-II Weyl points.}--- In contrast to Dirac points, Weyl points generically occur at low-symmetry $k$-points \cite{Soluyanov15p495}. Here, we find six pairs of Weyl points located in the $k_z=0$ plane and off the high-symmetry lines. To demonstrate the existence of Weyl points, and to pinpoint their location in $\bk$-space, we calculate the band structure on a dense $k$ grid on the $k_z = 0$ plane and show the energy difference between conduction and valence bands in Fig.\,\hyperref[{fig:fig_2}]{2(b)} on a logarithmic scale, {\it i.e.} $\log_{10}(E_{cb}-E_{vb})$. This clearly shows a pair of Weyl points $W_\pm$ symmetrically displaced from the $\Gamma\text{\textendash} K$ line. In Fig.\,\hyperref[{fig:fig_2}]{2(c)} we show the electronic bands from $\Gamma\text{\textendash}W_+\text{\textendash} K$, supporting the presence of an isolated touching of non-degenerate bands; the Weyl points have energy $E - E_{f} = + 0.25$\,eV relative to the Fermi energy. Contours of equal energy on the $k_{z}=0$ plane, corresponding to the energies of the red, black, and blue lines in (c), are presented in Fig.\,\hyperref[{fig:fig_2}]{2(d)}, suggesting that $W_\pm$ are type-II Weyl points. The Brillouin zone coordinates of one pair of Weyl points in $\bk$-space are $W_\pm = (0.217, \pm 0.036, 0)$ $\rm{\AA}^{-1}$. The five remaining pairs of Weyl points are related by sixfold symmetry, which, in particular, implies that all occur at the same energy. Fig.\,\hyperref[{fig:fig_3}]{3(a)} illustrates the positions of all twelve Weyl points and two Dirac points in the first Brillouin zone. Interestingly, the spatial separation between two adjacent Weyl points $\Delta W = |W_+ - W_-|$ is found to be 0.07 $\rm{\AA}^{-1}$, which is greater than the currently known largest value of 0.05 $\rm{\AA}^{-1}$ in the CuTlTe$_{2}$-type chalcoprite family of materials \cite{Jiawei16p226801}. The separation of the Weyl points in momentum space is determined by the strong SOC of the Pb $p$ orbitals and suggests that SrHgPb is an ideal material for studying the topological characteristics of Weyl semimetals, in particular the surface Fermi arcs, to which we return below.\par

Weyl points are sources or sinks of Berry curvature and can thus be viewed as momentum-space analogs of magnetic monopoles. The presence of such monopoles is the unambiguous proof of the existence and robustness of Weyl points. To provide such proof, we calculate the Berry curvature from first-principles calculations using $\Omega(\bk) = \nabla_{\bk}\times A(\bk)$, where $A(\bk) = \sum_{n=1}^{N}i\langle u_{n\bk}|\nabla_{\bk}|u_{n\bk} \rangle$ is the Berry connection \cite{Wang06p195118} with electron filling per unit cell $N$ and periodic part of the Bloch wavefunctions $u_{n\bk}$. The result is shown in Fig.\,\hyperref[{fig:fig_3}]{3(b)}, where the arrows denote the in-plane direction of the Berry curvature at each $k$ point (on the $k_{z}=0$ plane), and the presence of sources and sinks is clearly seen. The red and blue dots mark Weyl points with positive and negative chiralities, respectively, which is further confirmed by explicitly calculating the Chern number \footnote{See supplemental material for computational method, Chern number and $\mathbb{Z}_2$ topological invariant calculations for SrHgPb, the band structure of SrHgPb in $P6_{3}mmc$ space group, the derivation of the effective Hamiltonian, and the band structures  of other material candidates in hexagonal $ABC$ crystals. The computational method includes Refs. \cite{Tong66p1, Hohenberg64pB864, Kresse96p11169, Blochl94p17953,  Perdew96p3865, Koelling77p3017, Mostofi08p685, Sancho85p851,Wu2017,Alexey11p235401, Yu11p075119, Greschp17p075146}.}. \par

{\it Surface states.}---The topological nature of bulk semimetals is reflected in the structure of the electronic excitations on sample surfaces and boundaries. In particular, semimetals hosting Weyl fermions have special Fermi arc surface states on surfaces for which the bulk Weyl points with positive and negative chirality do not project onto the same point in the surface BZ. In the present case, this implies that Fermi arcs only appear on the (001) surface. To examine the surface states in detail, we calculated the (001) surface band structure of SrHgPb using a maximally-localized Wannier function (MLWF) Hamiltonian \cite{Mostofi08p685}. The surface spectral function, obtained through the  recursive Green{\textquoteright}s function method \cite{Sancho85p851}, is shown in Fig.\,\hyperref[{fig:fig_3}]{3(c)}. The spectral function for fixed energy $E - E_{f} = + 0.25$\,eV is presented in Fig.\,\hyperref[{fig:fig_3}]{3(d)}, clearly showing the Fermi arcs connecting bulk Weyl points of opposite chirality (indicated by red/blue dots). \par

{\it Phase transitions.}---To elucidate the existence of Weyl points, we have studied the electronic structure of SrHgPb as a function of the buckling of the honeycomb HgPb layers. In the limit of vanishing buckling, inversion symmetry is restored, and the space group is promoted to $P6_{3}/mmc$. This may be compared to other hexagonal $ABC$ ferroelectric materials \cite{Bennett12p167602}. The evolution of the total energy as a function of the buckling parameter $d$ is shown in Fig.\,\hyperref[{fig:fig_4}]{4(a)} and confirms that the buckled LiGaGe-type phase is lower in energy than the centrosymmetric phase. Interestingly, the band structure of SrHgPb calculated in the centrosymmetric $P6_{3}/mmc$ phase shows a full gap on the $k_z=0$ plane and no Weyl points \cite{Note1}. This implies that, as a function of increasing buckling, a topological transition occurs at which Weyl points are created. To study this transition, we have calculated the energy bands of a sequence of buckled structures, from $P6_{3}mc(+)$ to $P6_{3}mc(-)$ (see Fig.\,\hyperref[{fig:fig_4}]{4(a)}), and have tracked the presence and location of Weyl points. The result is presented in Fig.\,\hyperref[{fig:fig_4}]{4(b)}, and demonstrates that Weyl points are created at finite buckling strength $d_c \approx 0.06$ $\rm{\AA} $  on the mirror symmetric $\Gamma\text{\textendash}M$ line. As the buckling is further increased to $0.85$ $\rm{\AA}$, the Weyl points move towards the $\Gamma\text{\textendash}K$ line, on which they eventually annihilate in pairs as shown in Fig.\,\hyperref[{fig:fig_4}]{4(b)}. \par

These results indicate that the presence of the Weyl points in SrHgPb can be understood as an topological phase boundary between two gapped states in the $k_z=0$ plane. Notably, these two gapped time-reversal invariant $k_z=0$ subsystems differ in their topological $\mathbb{Z}_2$ index. Our Wilson loop calculations \cite{Alexey11p235401, Yu11p075119} reveal that the $k_z=0$ plane hosts nontrivial $\mathbb{Z}_2$ index in the centrosymmetric limit without buckling \cite{Note1}, while it has trivial $\mathbb{Z}_2$ index in the noncenstrosymmetric limit with strong buckling, where the Weyl points have been annihilated on the $\Gamma\text{\textendash} K $ line. This confirms that SrHgPb realizes an intermediate phase between distinct topological electronic states, and establishes a link with the general framework of topological phase transitions in non-centrosymmetric systems \cite{Murakami07p356,liu16p1663}. We expect that, by slightly breaking the rotational symmetry protecting the Dirac points,  the system becomes a strong topological (normal) insulator with $\mathbb{Z}_2$ topological indices (1;000) [(0;000)] in the centrosymmetric (noncentrosymmetric) phase.\par

{\it Model Hamiltonian.}---To obtain a qualitative understanding of the Dirac-Weyl phase, we consider a low-energy effective Hamiltonian around the $\Gamma$ point \cite{Note1}. Note the Dirac points shown in Fig.\,\hyperref[{fig:fig_2}]{2(a)} are not located in the vicinity of $\Gamma$, however, we find that a low-energy $k\cdot p$-type theory is capable of describing the essential physics of the Dirac-Weyl phase. We introduce a basis for the $\Gamma_7$ and $\Gamma_9$ doublets given by $\{|\Gamma_9,\tfrac{3}{2}\rangle, |\Gamma_7,\tfrac{1}{2}\rangle,|\Gamma_7,-\tfrac{1}{2}\rangle,|\Gamma_9,-\tfrac{3}{2}\rangle \}$ and find it convenient to write the Hamiltonian $H_\bk $, expanded up to cubic order in $\bk$, as a sum of terms $h^{\pm}_\bk$ which are even/odd ($+$/$-$) under inversion, i.e., $H_\bk = h^+_\bk+h^-_\bk$. The inversion symmetric part $h^+_\bk$ can then be expressed as
\be
h^+_\bk = 
\begin{pmatrix} m_\bk  & ig_1k_- & i g_3k_zk^2_- &  0 \\
-ig_1k_+& -m_\bk & 0  & -i g_3k_zk^2_- \\
-i g_3k_zk^2_+ &0 & -m_\bk &  ig_1k_-  \\
0 & i g_3k_zk^2_+ &-ig_1k_+ & m_\bk \\
 \end{pmatrix} ,  \label{eq:Heven}
\ee
where the mass $m_\bk=m_0 - m_1(k_x^2+k^2_y)-m_2k^2_z$; the coefficients $g_1,g_3$ are real parameters. The Hamiltonian $h^+_\bk$ is essentially equivalent to the low-energy $k \cdot p$ theory of inversion symmetric Dirac semimetals such as Na$_3$Bi \cite{Wang12p195320}, which have point groups isomorphic to $D_{6h}$. It describes two symmetry-protected Dirac points located at $k_z=\pm \sqrt{m_0/m_2}$, where $m_0$ corresponds to the band inversion at $\Gamma$.\par

The inversion asymmetric part of the Hamiltonian $h^-_\bk$, which lifts the double degeneracy of bands (except on the rotation axis), takes the form
\be
h^-_\bk = 
\begin{pmatrix} 0  & g_2k_zk_- & g'_2k^2_- & i\lambda_{\bk}  \\
g_2k_zk_+& 0 & -i g'_1k_-  & g'_2k^2_-  \\
g'_2k^2_+  &i g'_1k_+ & 0 &  -g_2k_zk_-  \\
-i\lambda^*_{\bk}  & g'_2k^2_+ & -g_2k_zk_+ & 0 \\
 \end{pmatrix}  ,    \label{eq:Hodd}
\ee
with the cubic terms given by $\lambda_{\bk}=g'_3k^3_++g''_3k^3_- $ (and all $g$-parameters real). The terms proportional to $g'_3$ reflect the true sixfold symmetry of the crystal; all other terms have an emergent full rotational symmetry in the $xy$-plane. The Dirac-Weyl semimetal phase can be accessed in the following way. For nonzero $g'_3$ and $g''_3$, as $g'_2$ is increased, Weyl points are created on one of the two sets of inequivalent vertical mirror planes. As $g'_2$ is further increased, the Weyl points traverse through the $k_z=0$ plane and annihilate in pairs on the other set of vertical mirror planes (for details see Supplemental Material \cite{Note1}). This is in full agreement with the numerical first-principles calculations.\par

\begin{figure}[pt]
    \centering
    \includegraphics[width=8.6cm]{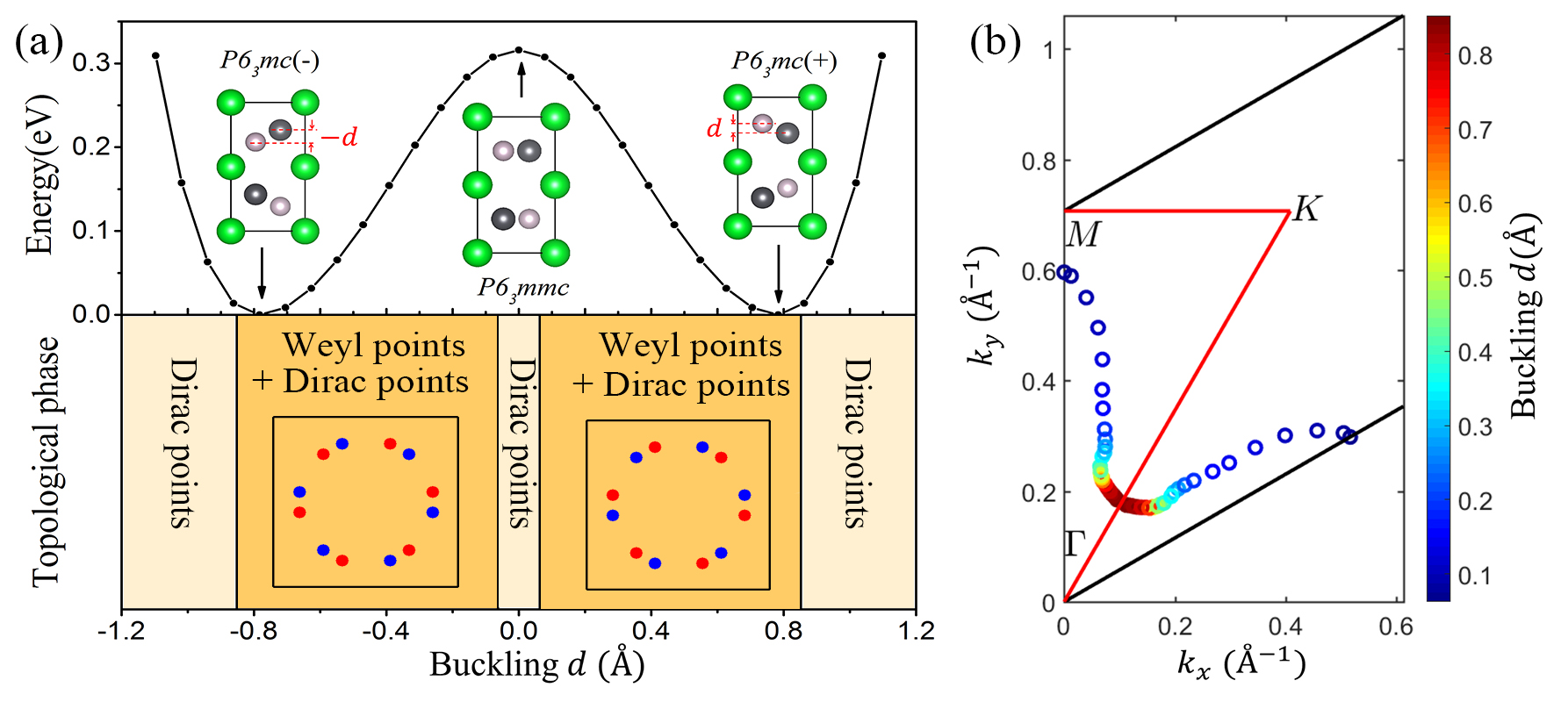}
    \caption{(a) The relative energy with respect to the buckling distance $ d$ of HgPb layer, below which we show the topological phase transitions due to the buckling
    $ d$, and the chirality of Weyl points. (b) The position evolution paths of the Weyl points are illustrated for various bucklings $ d$.}
    \label{fig:fig_4}
\end{figure}

{\it Other material candidates.}---Our first-principles results show that SrHgPb realizes the Dirac-Weyl semimetal phase. To expand the class of candidate material we have examined other experimentally synthesized hexagonal $ABC$ systems with LiGaGe-type crystal structure and group II-XII-IV elements. We find from first-principles calculations that SrHgSn and CaHgSn are also expected to realize the Dirac-Weyl semimetal phase \cite{Note1}. Furthermore, going beyond these documented materials, the ternary half-Heusler compounds are also promising candidates when transformed to the LiGaGe-type phase under pressure \cite{Xie14p6}. In particular, we have confirmed that LiGaGe-type CaAuBi \cite{Xie14p6} can host both Weyl points and Dirac points under 18\,GPa \cite{Note1}.\par

{\it Conclusion.}---Based on both first-principles calculations and a low-energy model Hamiltonian, we have proposed a new topological semimetal phase, characterized by the coexistence of Dirac points and Weyl points in momentum space. The coexistence of these two exotic quantum matter states can be realized in a known material LiGaGe-type hexagonal lattice SrHgPb. The topological phase transition driven by HgPb buckling in SrHgPb reveals that the position and chirality of Weyl points can be tuned by the buckling, and the Weyl points will not annihilate unless two Weyl points with  opposite chirality touch each other. Other possible material realizations of LiGaGe-type in experiment are discussed. The coexistence of Dirac points and Weyl points in $ABC$  hexagonal crystals will provide an interesting platform to study the interplay between Dirac and Weyl points, and for understanding their transport and optical properties.\par

\begin{acknowledgments}
This work was supported by National Key Basic Research Program of China (Grant No. 2015CB921600), the National Natural Science Foundation of China (Grants No. 51672171), the Eastern Scholar Program from the Shanghai Municipal Education Commission, and the fund of the State Key Laboratory of Solidification Processing in NWPU (SKLSP201703). Special Program for Applied Research on Super Computation of the NSFC-Guangdong Joint Fund (the second phase), the supercomputing services from AM-HPC, and the Fok Ying Tung Education Foundation are also acknowledged. Y.K. acknowledges support from National Science Foundation under Grant No. DMR-1120901 and the National Research Foundation of Korea (NRF) grant funded by the Korea government (MSIP; Ministry of Science, ICT $\&$ Future Planning) (No. S-2017-0661-000). C.L.K. acknowledges support from a Simons Investigator grant from the Simons Foundation. E.J.M.'s work on this project was supported by the U.S. Department of Energy, Office of Basic Energy Sciences under Award No. DE-FG02-84ER45118. A.M.R. acknowledges support from the DOE Office of Basic Energy Sciences, under Grant No. DE-FG02-07ER46431. H.G. acknowledges the support of China Scholarship Council.
\end{acknowledgments}

\bibliography{ref}

\widetext
\clearpage


\section*{Supplementary material}

# Supplementary Material of "The Dirac-Weyl semimetal: Coexistence of Dirac and Weyl fermions in polar hexagonal $ABC$ crystals"

Heng Gao,[1,2] Youngkuk Kim,[2,3] Jörn W. F. Venderbos,[2,4]
C. L. Kane,[4] E. J. Mele,[4] Andrew M. Rappe,[2,*] and Wei Ren[1,†]

[1]*International Centre for Quantum and Molecular Structures, Department of Physics, Shanghai University, 99 Shangda Road, Shanghai 200444, China*

[2]*Department of Chemistry, University of Pennsylvania, Philadelphia, Pennsylvania 19104–6323, USA*

[3]*Department of Physics, Sungkyunkwan University, Suwon 440–746, Korea*

[4]*Department of Physics and Astronomy, University of Pennsylvania, Philadelphia, Pennsylvania 19104–6396, USA*

## A.  Computational method

In order to demonstrate the coexistence of Weyl and Dirac points in $ABC$ crystals, we performed the structural optimization and electronic structures calculations within the framework of density functional theory (DFT)[1, 2]. We used Vienna *ab initio* simulation package (VASP) [3] based on the projector augmented wave (PAW) method [4]. The exchange-correlation interaction was treated within the generalized gradient approximation (GGA) [5] parametrized by Perdew, Burke, and Ernzerhof (PBE). The energy cutoff of 500 eV was set in all the calculations and a Monkhorst-Pack grid with $11 \times 11 \times 7$ k-points was used for Brillouin zone integration. For the structural optimization, the lattice parameters and all the atoms are assumed to be relaxed when the Hellmann-Feynman forces on all atoms are less than $0.01\,\text{eV/Å}$. Spin-orbit coupling (SOC) was taken into account self-consistently in terms of the second vibrational procedure [6]. To explore the surface states and Fermi arc of the SrHgPb, the Hamiltonian was constructed by using maximally-localized Wannier function (MLWF) for $p$ orbitals of Pb and $s$ orbitals of Hg as implemented in Wannier90 package [7]. The surface states and Fermi arc have been calculated using recursive Green's function method [8] based on MLWF using the WannierTools [9]. The Chern number and $\mathbb{Z}_2$ topological invariant are calculated by the Wannier charge center using Wilson loop method [10, 11] as implemented in Z2pack [12].

## B.  Topological invariant of SrHgPb in $P6_3mc$ space group

In this section, we use the Wilson loop method to calculate the Chern number and $\mathbb{Z}_2$ topological invariant of SrHgPb in $P6_3mc$ space group. For the Chern number calculations, a closed sphere is chosen, such that it encloses a single Weyl point. The sphere is parametrized by the discretized angles $\theta$ and $\varphi$ ($0 \leq \theta \leq \pi$, $0 \leq \varphi \leq 2\pi$). The Chern number is calculated by the Berry phase, corresponding to the average position of charge associated with the occupied bands on the loop $\varphi$ $\theta_i$ [13]. The results for a pair of Weyl points with opposite chirality are shown in Fig. S1(a). The calculated Chern number for a pair of Weyl points are +1 and -1, respectively. The $\mathbb{Z}_2$ topological invariant for the $k_z = \pi$ plane is shown in Fig. S1(b). The Wannier charge center flows (Wilton loops) cross the dashed reference line (colored by red) even times, which results in trivial topological invariant in the $k_z = \pi$ plane.

---

* rappe@sas.upenn.edu
† renwei@shu.edu.cn

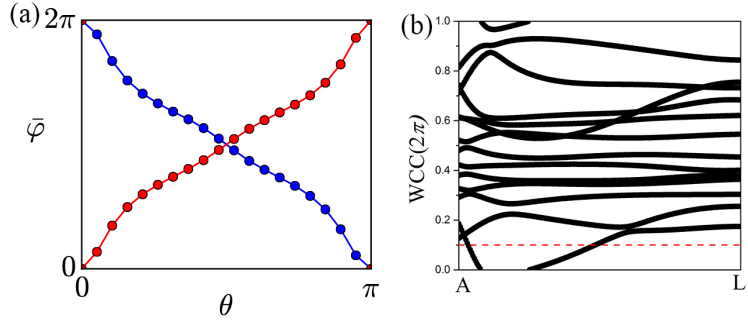

FIG. S1. (a) The evolution of the sum of Wannier Charge center (WCC) on the sphere that encloses the Weyl point. The red and blue curves show the evolution of for Weyl points with the +1 and -1 Chern numbers, respectively. (b) The Wannier center charge (WCC) evolution (Wilson loop) along the $A - L$ line on the $k_z = \pi$ plane. The red dashed line, crossing with the WCC even times, is referenced to evaluate the trivial $\mathbb{Z}_2$ topological invariant.

### C. Band structure and $\mathbb{Z}_2$ invariant of SrHgPb in $P6_3/mmc$ space group

In this section, we use the Wilson loop method to calculate the $\mathbb{Z}_2$ topological invariant of SrHgPb in the centrosymmetric $P6_3/mmc$ phase. The GGA band structure of SrHgPb in $P6_3/mmc$ space group is shown in Fig. S2(a). The Dirac points along $\Gamma - A$ still survive, while the Weyl points disappear in the centrosymmetric phase, in accordance with the fact that the Weyl points are not allowed in centrosymmetric and time-reversal symmetric system. The WCC of SrHgPb in the $k_z = 0$ plane, shown in Fig. S2(b), suggests that the $k_z = 0$ plane is topologically nontrivial in the centrosymmetric phase.

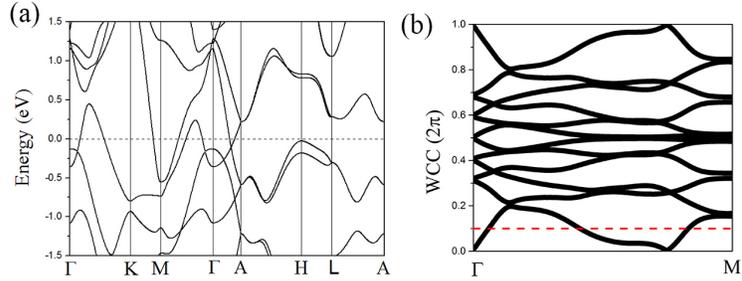

FIG. S2. (a) The band structure of SrHgPb in $P6_3/mmc$ space group (b) The WCC evolution along the $\Gamma - M$ line at $k_z = 0$ plane. The WCC crosses the reference line three times, indicating that the $k_z = 0$ plane hosts nontrivial $\mathbb{Z}_2$ topological invariant.

### D. Derivation of the effective Hamiltonian

We construct an effective $k \cdot p$ model Hamiltonian, describing the conduction $\Gamma_7$ state and valence $\Gamma_9$ state near the $\Gamma$ point. The Hamiltonian is constructed such that it respects the little group of $\Gamma$ points, which is $C_{6v}$, together with the time reversal symmetry $\mathcal{T}$. The symmetry generators of $C_{6v}$ include 6-fold rotation $C_{6z}(\frac{x}{2} - \frac{\sqrt{3}}{2}, \frac{\sqrt{3}}{2}x + \frac{y}{2}, z)$ and mirror symmetry $M_{xz}(x, -y, z)$. The effective Hamiltonian is constructed under the constraint

$$D(g)H(\mathbf{k})D^\dagger(g) = H(g\mathbf{k}), \tag{1}$$

for a symmetry operation $g$, and its representation $\mathcal{D}(g)$. The representations of above symmetry operators in

the basis $|\Gamma_9, \frac{3}{2}\rangle$, $|\Gamma_7, \frac{1}{2}\rangle$, $|\Gamma_7, -\frac{1}{2}\rangle$, $|\Gamma_9, -\frac{3}{2}\rangle$ are given by

$$D(C_{6z}) = \text{diag}\{-i, \exp(-\frac{i\pi}{6}), \exp(\frac{i\pi}{6}), i\} \tag{2}$$

$$D(M_{xz}) = \begin{pmatrix} 0 & 0 & 0 & -1 \\ 0 & 0 & 1 & 0 \\ 0 & -1 & 0 & 0 \\ 1 & 0 & 0 & 0 \end{pmatrix}. \tag{3}$$

$$D(\mathcal{T}) = \begin{pmatrix} 0 & 0 & 0 & 1 \\ 0 & 0 & -1 & 0 \\ 0 & 1 & 0 & 0 \\ -1 & 0 & 0 & 0 \end{pmatrix}. \tag{4}$$

One can obtain the effective Hamiltonian up to cubic terms in $\mathbf{k}$ as

$$H_{\mathbf{k}} = M_{\mathbf{k}} + \begin{pmatrix} m_{\mathbf{k}} & ig_1 k_- + g_2 k_z k_- & ig_3 k_z k_-^2 + g_2' k_+^2 & i\lambda_{\mathbf{k}} \\ -ig_1 k_+ + g_2 k_z k_+ & -m_{\mathbf{k}} & -ig_1' k_- & -ig_3 k_z k_-^2 + g_2' k_-^2 \\ -ig_3 k_z k_+^2 + g_2' k_+^2 & ig_1' k_+ & -m_{\mathbf{k}} & ig_1 k_- - g_2 k_z k_- \\ -i\lambda_{\mathbf{k}}^* & ig_3 k_z k_+^2 + g_2' k_+^2 & -ig_1 k_+ - g_2 k_z k_+ & m_{\mathbf{k}} \end{pmatrix}, \tag{5}$$

where $M_k = M_0 + M_1(k_x^2 + k_y^2) + M_2 k_z^2$, $m_k = m_0 - m_1(k_x^2 + k_y^2) - m_2 k_z^2$, $\lambda_{\mathbf{k}} = g_3' k_+^3 + g_3'' k_-^3$, and $k_\pm = k_x \pm k_y$. To demonstrate this low-energy effective Hamiltonian can capture both Dirac and Weyl points, the bands structures and the energy difference between conduction and valence bands in the $k_z = 0$ plane are shown in Fig. S3. The band structures and the pattern of Weyl points in the $k_z = 0$ plane agree with our first principles calculations. The creation and annihilation of Weyl points in two sets of inequivalent the vertical mirror planes $k_x = 0$ and $k_y = 0$, as well as the position of Weyl points in the $k_z = 0$ plane can be controlled via $g_2''$, $g_3'$, and $g_3''$ parameters.

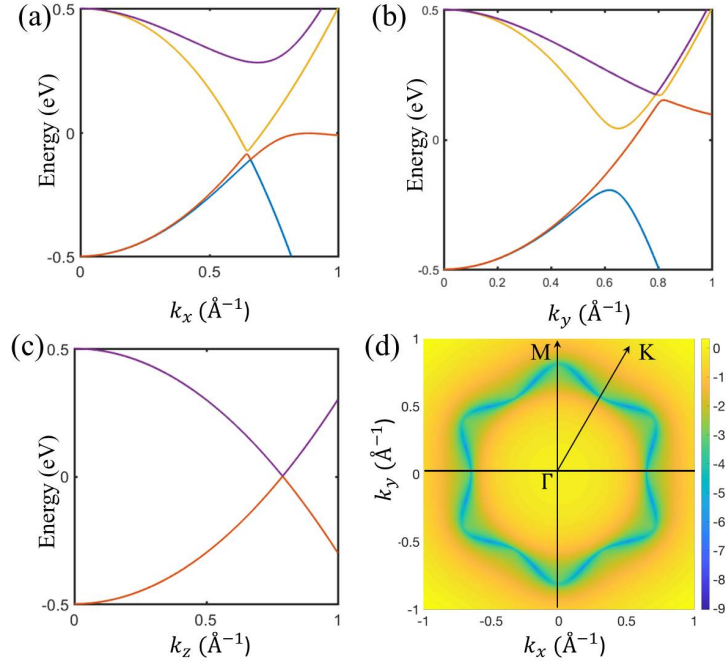

FIG. S3. The band structures along (a) $\Gamma - K$ (b) $\Gamma - M$ (c) $\Gamma - A$ lines calculated from the $k \cdot p$ model Hamiltonian. (d) The contour of energy difference between the conduction and valence bands in logarithmic scale $\log_{10}(E_{cb} - E_{vb})$ in the $k_z = 0$ plane. Here we used the parameter set, $M_k = 0$, $m_0 = 0.5$, $m_1 = 1$, $m_2 = 0.8$, $g_1 = 0.1$, $g_1' = 0$, $g_2 = 0.1$, $g_3 = 0.1$, $g_2' = 0.14$, $g_3' = 0.6$, and $g_3'' = 0$.

### E. Weyl nodal line phase and Dirac points in SrHgPb

In this section, we discuss the Dirac points and coexisting Weyl nodal line (WNL) in SrHgPb, and their evolution as the buckling strength changes. Figure S4(a) shows the band structure of SrHgPb with buckling of 0.89 Å. We note that the band crossing points between the conduction and valence bands occur along $\Gamma - K$ line. This is an indicative of the WNL resinding in the vertical mirror plane containing $\Gamma$, $K$, $H$, and $A$. From the energy difference between the conduction and valence bands in the $k_y = 0$ plane, shown in the Fig. S4(d), the position of WNL can be identified, which corrersponds to the bluish lines in the contour plots. We find that the size and position of WNL can be tuned via the buckling strength. The WNL disappears when the buckling is larger than 1.5 Å. The conduction and valence bands in the $k_y = 0$ plane have opposite mirror eigenvalues, Therefore, the existence of WNL can be considered as being protected by the vertical mirror symmetry. More interestingly, we identify the coexisting double Dirac points at $A$, which is a composite of two Dirac points, at the buckling of 1.01 Å. The conduction and valence bands meet at the $A$ point at the critical buckling strength. Figure S4(b) shows the band structure of the eight-fold degenerate double Dirac point at the critical value of buckling. This double Dirac point is not a symmetry-guaranteed, since the change of buckling strength away from the critical buckling strength, which can be done preserving the crystalline symmetries, would lift this eight-fold degeneracy. When the buckling of SrHgPb is greater than the critical value of 1.01 Å, the double Dirac points split into two Dirac points after the band inversion between two conduction bands and two valence bands at $A$ point. We find that the existence of Dirac points is robust against the perturbation changing the buckling strength, and the Dirac points survive upto the buckling of 1.71 Å. Figure S4(c) shows the band structure of SrHgPb with buckling strength of 1.71 Å. The Dirac points appear on the $\Gamma - A$ line.

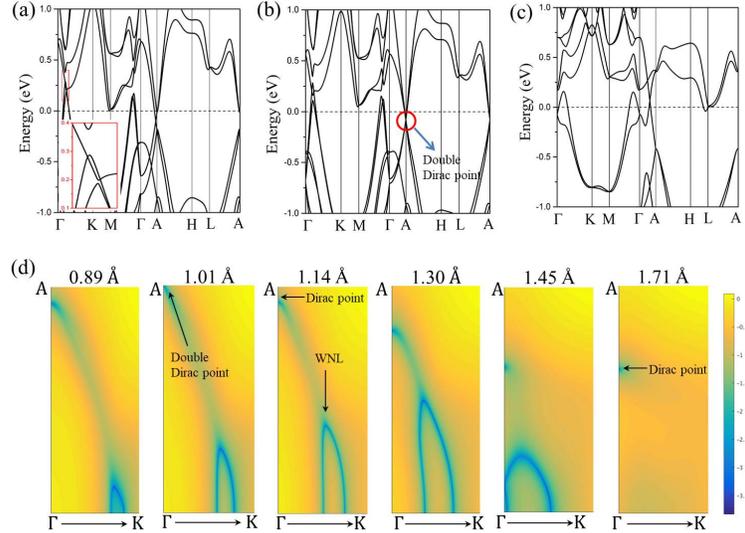

FIG. S4. The band structures of SrHgPb with the buckling strengths of (a) 0.89Å, (b) 1.01Å, and (c) 1.71Å. (d) The contour of the energy difference between the conduction and valence bands in logarithmic scale $\log_{10}(E_{cb} - E_{vb})$ in the $k_y = 0$ plane of SrHgPb with a of various values of buckling strength, which are presented above the figures.

### F. Other material candidates in hexagonal $ABC$ crystals

In this section, we show that other materials in the hexagonal $ABC$-type crystals, such as SrHgSn, CaHgSn, and CaAuBi, can host both the coexisting Dirac and Weyl points in momentum space. Encouragingly, SrHgSn and CaHgSn are existing compounds, synthesized in the experiment [14]. CaAuBi in $P6_3mc$ space group can be transformed from half-Heusler ternary phase under pressure of 18 GPa [15]. The band structures of these three material candidates have been calculated, and shown in Fig. S5. They all feature Dirac point along $\Gamma - A$. To prove the coexistence of Weyl points in $k_z = 0$ plane, we calculated the energy difference between conduction and valence bands in $k_z = 0$ plane of material candidates, which is shown in Fig. S5. These figures clearly show a pair of Weyl points off a high-symmetric position near the $\Gamma - K$ line.

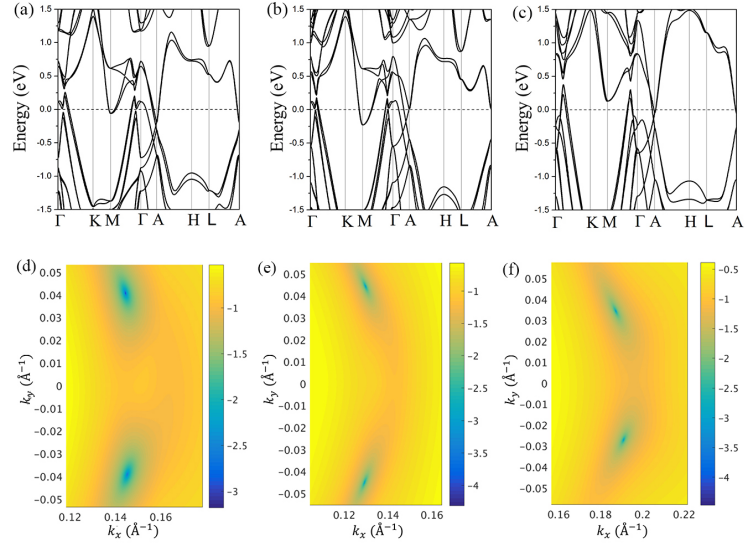

FIG. S5. The band structures of (a) SrHgSn, (b) CaHgSn, and (c) CaAuBi. The contour of energy difference between the conduction and valence bands in logarithmic scale $\log_{10}(E_{cb} - E_{vb})$ in the $k_z = 0$ plane for (d) SrHgSn, (e) CaHgSn, and (f) CaAuBi. The blue points indicate the Weyl points.

---



\end{document}